\documentclass[12pt,a4paper,dvips]{article}
\usepackage{ a4p}
\usepackage{cite,mcite,epsfig,rotating}
\usepackage{graphicx}
\usepackage{physics }
\usepackage{l3_title,ifthen, Lep}
\date{June 28, 2001}
\lthreenote{XXXX}
\preprint{2001-046}
\newcommand{\rar}{\rightarrow}

\def\mm{\mbox{$M \hskip 0.010in \! \! \! \! \! / _{ll}$}}

\def\Nl{\ifmmode \mathrm{N_\ell} \else
                $\mathrm{N_\ell}$   \fi}%
\def\Ne{\ifmmode \mathrm{N}_e \else
                $\mathrm{N}_e$   \fi}%
\def\Nm{\ifmmode \mathrm{N}_\mu \else
                $\mathrm{N}_\mu$   \fi}%
\def\Nt{\ifmmode \mathrm{N}_\tau \else
                $\mathrm{N}_\tau$   \fi}%
\def\Ul{\ifmmode U_{\ell N} \else
                $U_{\ell N}$   \fi}%
\def\Ue{\ifmmode U_{eN} \else
                $U_{eN}$   \fi}%
\def\MZ{m_{\mathrm{Z}}}
\def\MW{m_{\mathrm{W}}}

\def\Ne{\ifmmode \mathrm{N}_e\else
                $\mathrm{N}_e$\fi}%
\def\Nm{\ifmmode \mathrm{N}_\mu\else
                $\mathrm{N}_\mu$\fi}%
\def\Nt{\ifmmode \mathrm{N}_\tau\else
                $\mathrm{N}_\tau$\fi}%

\newlength{\capindent}
\newlength{\capwidth}
\newlength{\figwidth}
\newcommand{\icaption}[2][!*!,!]{\hspace*{\capindent}%
  \begin{minipage}{\capwidth}
    \ifthenelse{\equal{#1}{!*!,!}}%
      {\caption{#2}}%
      {\caption[#1]{#2}}
  \end{minipage}}
\begin{document}
\bibliographystyle{l3style}
\begin{titlepage}
 \title{ Search for Heavy Neutral 
  and Charged Leptons \\ 
 in e$^+$e$^-$ Annihilation at LEP }
\author{The L3 Collaboration}
%
%
\begin{abstract}
A search for exotic unstable neutral
and charged heavy leptons as well as for  
stable charged heavy leptons is performed 
with the L3 detector at LEP. Sequential,
vector and mirror natures of heavy leptons are considered. 
No evidence for their 
existence is found and lower limits on their masses are set. 
\end{abstract}

%
%
\submitted
 \end{titlepage}

\section*{Introduction}

Many theories beyond the Standard Model 
of the electro-weak interation~\cite{hlep00} 
predict the existence of new heavy leptons~\cite{hlep01} 
with observable production cross sections and cleanly 
identifiable final states.
Results on this subject obtained 
at the Z resonance by LEP and SLC experiments 
are found in Reference~\citen{hlep02}.  
Results obtained by the LEP experiments
at centre-of-mass energies, $\sqrt{s}$ = 133$-$189 \GeV\, are found 
in References~\citen{hlep02a}~and~\citen{hlep03}.
Here we report on a search for exotic pair-produced heavy leptons. 
The data used in this analysis were collected 
with the L3 detector~\cite{hlep04}
at LEP during 1999 at $\sqrt{s}$ = 192$-$202 \GeV\ and 2000 at 
$\sqrt{s}$ = 200$-$208 \GeV\ with a total 
integrated luminosity 450 pb$^{-1}$. 
The results are combined with 
our earlier data recorded at $\sqrt{s}~=~133-189$ \GeV.

The exotic leptons discussed in this paper may be classified by
their $SU(2) \times U(1)$ quantum numbers~\cite{hlep01}:

\begin{itemize}

\item Sequential leptons: they exist in the simplest extension of
          the Standard Model where a fourth family with the same 
          quantum numbers is added to the known fermionic spectrum. 
          The indirect limits on existence of these particles 
          can be found in Reference~\citen{hlep041}.

\item Vector leptons: these particles occur in both left-handed and
           right-handed weak isodoublets. 
           Among others, the E$_6$ group model~\cite{hlep042} predicts them.  

\item Mirror leptons~\cite{hlep043}: these particles have chiral 
            properties which are opposite to those of ordinary leptons,
            {\it i.e.} the right-handed components form a weak isodoublet and
            the left-handed ones are weak isosinglets. Mirror fermions
            appear in many extensions of the Standard Model 
            and provide a possible  way to restore left-right 
            symmetry at the scale of the
            electroweak symmetry breaking.

\end{itemize}

\section*{Production and decay of exotic leptons}

Charged heavy leptons, $\rm L^\pm$, are pair-produced through the 
$s$-channel via $\gamma$ and Z boson exchange, 
while for heavy neutral leptons, $\rm L^0$, only 
the Z boson exchange is present.
The total cross sections are in the range of 1--3 pb at masses well 
below the beam energy, $E_{\it beam}$, and fall as the mass of the 
lepton approaches the beam energy. 
The neutral heavy leptons are either of Dirac or Majorana types. 
The main difference between pair-produced
Dirac and Majorana neutral heavy leptons is the dependence of the cross section
on their velocity $\beta$.
A $\beta (3-\beta^2)$ term is present in the Dirac case while the cross
section in the Majorana case is proportional to $\beta^3$.
This implies that the Majorana cross section 
falls more rapidly with mass than the Dirac one. 
In this search, heavy leptons couple
to the electron, muon, or tau families.
Only the case where both heavy leptons decay into the same light
lepton family is considered.

Exotic leptons decay via the charged or neutral weak currents
through the mixing with a light lepton, except for sequential  leptons,
which decay only through charged current:

\vskip 0.3cm
\centerline{
L$^\pm 
 \rightarrow \nu_{\ell}  {\rm W^\pm} $ 
 or
L$^\pm \rightarrow \ell^\pm$Z
 }
\vskip 0.3cm
\centerline{
L$^0 
 \rightarrow \ell^\pm {\rm W^\mp} $ 
 or
L$^0 \rightarrow \nu_{\ell}$Z 
 }
\vskip 0.3cm

\noindent
where $\ell = {\rm e}, \mu, \tau$.
For masses, $m_{\mathrm{L}}$, below 100 \GeV, 
heavy leptons decay predominantly through the W boson.
The branching fraction of 
$\rm L^0 \rightarrow eW$ is about 90\% for $m_{\mathrm{L}}$ = 100 \GeV\
and 100\% for $m_{\mathrm{L}}$ $\leq$ 90 \GeV.
Therefore, only the decay into the W boson is considered.

The decay amplitude of L$^0 \rightarrow \ell^\pm {\rm W^\mp} $ 
 contains a mixing parameter $U$.  
The mean decay length, $D$, as a function of 
$\vert U \vert {}^2$ and mass is given by
$ D = \beta \gamma c \tau_{\mathrm{L}}  \propto \beta
 \vert U \vert^{-2} m_{\mathrm{L}}^{\alpha}$~\cite{hlep055},
where $\tau_{\mathrm{L}}$ is the lifetime of the 
heavy lepton and $\alpha \approx -6$. 
This implies the decay can occur far
from the interaction point if the particle has a low mass 
or a very small coupling.
To ensure high detection and reconstruction efficiencies, 
this search is
restricted to decay lengths below 1cm, which
limits the sensitivity to the mixing parameter to
$\vert U \vert {}^2$ $>$ $\mathcal{O}($10$^{-11})$.

Two additional possibilities for the charged heavy lepton decays
are considered:

\begin{itemize}
\item The charged leptons decay through a weak charged current interaction
      that conserves lepton number, L$^{\pm} \rar $L$^0 $W$^{\pm}$.
\item The charged leptons are stable. This is the case 
         if the associated neutral lepton is heavier than its
         charged partner and the coupling with light leptons is 
         negligible.
\end{itemize}

\section*{Event simulation}

The generation of heavy leptons and their decay is performed with 
the TIPTOP~\cite{hlep06} Monte Carlo program which 
takes into account initial state radiation and spin effects.
A mass range between 50 and 100 \GeV\
for the heavy leptons is considered.
For the simulation of background from Standard Model processes the
following Monte Carlo programs are used:
KK2f~\cite{hlep060} ($\mbox{e}^+ \mbox{e}^- \rar \mbox{q} \bar{\mbox{q}} (\gamma)$),
PYTHIA~\cite{hlep07} ($\mbox{Z} \mbox{e}^+ \mbox{e}^-,~\mbox{Z} \mbox{Z}$),
BHWIDE~\cite{hlep070} ($\mbox{e}^+ \mbox{e}^- \rar \mbox{e}^+ \mbox{e}^- (\gamma)$),
KORALZ~\cite{hlep071} ($\mbox{e}^+ \mbox{e}^- \rar \mu^+ \mu^- (\gamma)$
                   and $\mbox{e}^+ \mbox{e}^- \rar \tau^+ \tau^- (\gamma)$),
KORALW~\cite{hlep072} ($\mbox{e}^+ \mbox{e}^- \rar \mbox{W}^+ \mbox{W}^- $),
PHOJET~\cite{hlep08} ($\mbox{e}^+ \mbox{e}^- \rar \mbox{e}^+ \mbox{e}^- \mbox{q} \bar{\mbox{q}}$),
DIAG36~\cite{hlep081} ($\mbox{e}^+ \mbox{e}^- \rar \mbox{e}^+ \mbox{e}^- \ell^+ \ell^-$),
and EXCALIBUR~\cite{hlep082} for other four-fermion final states. 
The Monte Carlo events are 
simulated in the L3 detector using the GEANT3 program~\cite{hlep09}, 
which takes into account the effects of energy loss, 
multiple scattering and showering in the materials. 
Time dependent detector inefficiencies, as monitored during the data
taking period, are also simulated.

\section*{Search for unstable heavy leptons}

\subsection*{Lepton and jet identification}

Unstable heavy lepton signatures are characterized by hadronic 
jets, isolated leptons, and missing energy and transverse momentum. 

An electron is identified as a cluster in the electromagnetic 
calorimeter with an energy larger than 4 \GeV\ matched to a track 
in the $(R,\phi)$ plane to within 20 mrad. 
The cluster profile should be consistent with that  expected for an electron. 
The polar angle of the electron candidate must satisfy $|\cos \theta|<0.94$.
A muon track is required to have  
track segments in at least two out of three layers of muon chambers
and to point back to the interaction region.
The muon should have a momentum greater than 4 \GeV\ and  
a polar angle defined by $|\cos \theta|<0.8$.
The lepton is isolated if 
less than
5 GeV is deposited in a 30$^{\circ}$ cone around its direction.
Moreover, to eliminate photon conversions, 
only one matched track is required for isolated electrons.
Jets are reconstructed from electromagnetic and hadronic calorimeter 
clusters using the Durham algorithm \cite{hlep091} with a
jet resolution parameter $y_{cut}$ = 0.008. The jet momenta are defined
by the vectorial energy sum of calorimetric clusters.

 The most important backgrounds 
are W$^+$W$^-$, $\mbox{q} \bar{\mbox{q}} (\gamma)$ and ZZ production. 
The W$^+$W$^-$ process is rejected by
cuts on visible mass, number of hadronic jets, jet angles
and energies and lepton energies. 
The $\mbox{q} \bar{\mbox{q}} (\gamma)$ process is suppressed by demanding 
transverse momentum imbalance, no energy in the very 
forward-backward regions, and a missing momentum vector pointing 
into the detector.
For ZZ event rejection, a cut on the energy sum of the two
isolated leptons is applied.

\subsection*{Search for unstable neutral heavy leptons}

The search for pair-produced neutral heavy
leptons requires two isolated leptons (e, $\mu$, or $\tau$) of the same 
family together with the decay products of W bosons,
$\it i.e.$  e$^+ $e$^- \rar $L$^0\overline{\mathrm{L}}^0
   \rar \ell^+\ell^- $W$^{+}$W$^{-}$.
Hadronic events with visible energy greater than 60 \GeV\ and
charged track multiplicity greater than 3 are considered.
 
For the case where both neutral heavy leptons decay to either
electrons or muons, 
events must also satisfy the following criteria:

\begin{itemize}

\item The number of reconstructed jets plus isolated leptons is 
      at least 3.

\item The event contains at least two isolated electrons or two isolated
      muons. Figure 1 shows the energy in a 30$^{\circ}$ cone for the 
      second most energetic electron candidate.

\item The energy sum of the two isolated electrons or muons 
      must be less than 0.7$\times E_{beam}$.
      Figure 2 shows the sum of the energies of two muons.
 
\end{itemize}

\noindent
After this selection, 11 events remain in 
the data for the electron decay mode while $ 10.7 \pm 0.5$ background events are expected.
For the muon decay mode, 5 events remain in the data while
$2.3 \pm 0.2$ background events are expected.  

For the case where both  neutral heavy leptons decay to tau leptons,
each of the taus decays independently into hadrons, muons, or electrons.
When both tau leptons decay into either muons or electrons, the above 
selection is applied with the exception that we allow the isolated leptons 
to be either two electrons, two muons, or one muon and one electron.
For the case in which at least one of the 
tau decays into one charged hadron, events are required to
satisfy the the following criteria: 

\begin{itemize}

\item The number of reconstructed jets plus isolated leptons 
          is at least 4.

\item The polar angle  
          of the missing momentum  must be in the range
          $25^{\circ}<\theta_{miss}<155^{\circ}$, as depicted in Figure 3, 
          and the fraction of the visible energy in the 
          forward-backward region,
          $\theta<20^{\circ}$ and $\theta>160^{\circ}$, must
          be less than 40\%.

\item The angle between the most isolated track and the track nearest to it
          must be greater than 50$^{\circ}$, or the angle 
          between the second most isolated track and the track nearest to it
          must be greater than 25$^{\circ}$.
          The transverse momenta of the two most isolated tracks 
          must be greater than 1.2 \GeV, 
          and at least one track must have a 
          transverse momentum  greater than 2.5~\GeV. 

\item The electron and muon energies must be less than 50 \GeV. 

\end{itemize}

\noindent
After applying the above selection and also accounting for the 
case where both taus decay into electrons or muons, 114 events 
remain in the data, while $121.5 \pm 1.6$ background events are expected.  
         
\subsection*{Search for unstable charged heavy leptons }

\subsubsection*{Decay into a light neutrino, 
L$^{\pm} \rar \nu_{\ell} $W$^{\pm}$}

For this search, two sets of cuts are used to identify the topology
with one hadronic and one leptonic W-decay as well as two hadronic
W-decays. For both selections, the visible energy 
is required to be greater 
than 50 \GeV, the multiplicity of charged tracks to be greater than 3, 
and the fraction of the total
visible energy in the forward-backward region to be less than 25\%.

For the decay L$^+$L$^-$ $\rightarrow$ $\nu_{\ell} \bar\nu_{\ell}$
W$^{+}$W$^{-}$ 
$\rightarrow$ $\nu_{\ell} \bar\nu_{\ell} \ell \nu_{\ell} 
\mbox{q} {\bar{\mbox{q}}}'$,
events satisfying the following criteria are selected:

\begin{itemize}

\item The event contains at least one isolated electron or muon with
          energy greater than 15~\GeV\ and less than 60 \GeV.

\item The number of reconstructed jets plus isolated leptons is at least 3. 

\item The polar angle of the missing momentum must be in the range
          $25^{\circ}<\theta_{miss}<155^{\circ}$.

\item     The sum of the energies of the hadronic jets must be less than
          0.85$\times E_{beam}$.
          Figure 4a shows the distribution of the sum of 
          the energies of hadronic jets.

\end{itemize}

\noindent
After applying the selection, 
76 events remain in the data while $ 73.1 \pm 1.4$ background 
events are expected.

For the mode where both W bosons decay hadronically, events satisfying 
the following criteria are selected:

\begin{itemize}

\item The event does not contain any isolated electron or muon and
      the energy of each non-isolated electron or muon 
      must be less than 30 \GeV. 

\item The number of hadronic jets is at least 4.

\end{itemize}

After application of the previous cuts,
the dominant remaining background is W-pair production. 
However, for the signal events, 
the total visible energy is less than
$\sqrt{s}$ and the two W bosons are not back-to-back
due to the energy and momentum carried
away by light neutrinos.
The following additional requirements are then imposed:

\begin{itemize}

\item The visible energy must be less than 0.9$\times$$\sqrt{s}$, as
      shown in Figure 4b.

\item The acollinearity angle between the 
      two W candidates is greater than 15$^{\circ}$,
      as shown in Figure 4c.

\item The total transverse momentum of jets must be greater than 10 \GeV,
      as shown in Figure 4d.

\end{itemize}

For charged heavy lepton masses greater 
than the W boson mass, the leptons decay to a real W.
Therefore, for the case where the charged lepton mass is greater than 80 \GeV,
a kinematic fit is applied 
to improve the determination of jet energies and angles.
Both jet-jet invariant masses are constrained to the W mass. 
All possible jet-jet combinations 
are considered and the one which yields 
the smallest $\chi^2$ of the fit is chosen.

After applying the selection, 79 events remain in the data while 
$75.0 \pm 1.5$ background events are expected.

\subsubsection*{Decay into a stable neutral heavy lepton, L$^{\pm} \rar $L$^0 $W$^{\pm}$}

The search for a charged heavy lepton decaying into a stable neutral
heavy lepton assumes a mass of the associated heavy neutrino L$^0$ to be 
greater than 40 \GeV. This assumption is based on LEP 
results at the Z resonance~\cite{hlep02} and implies 
a large missing energy and large transverse momentum imbalance. 
In the limit of a vanishing mass
difference between the charged lepton and its associated neutral lepton,
$\Delta m = m_{\rm L^\pm} - m_{\rm L^0}$, the signal efficiency
is affected by the trigger efficiency and the rejection of the 
two-photon background. Here the search is restricted to 
5 \GeV\ $\leq$ $\Delta m$ $\leq$ 60 \GeV.
The case of a light neutrino, $\Delta m$ = $m_{L^{\pm}}$, is considered
in the previous section.
The main background is the two-photon process for small mass
differences, $\Delta m~\leq$ 20 \GeV and the 
$\mbox{q} \bar{\mbox{q}} (\gamma)$ and W$^+$W$^-$ processes
for high mass differences, $\Delta m~\geq$ 20 \GeV.

This signature is very similar to that of a chargino 
decaying into a stable neutralino and a W boson. 
Therefore, a selection developed for the chargino search~\cite{charg} is used.
It is based on the signatures of missing energy,
transverse momentum imbalance and isolated leptons. Cuts on
missing mass and acoplanarity are also used. The selection is
optimized for three different $\Delta m$ regions: 
for very low $\Delta m$ region around 5 \GeV,
the low $\Delta m$ range from 10 to 30 \GeV\ and 
the medium $\Delta m$ range from 40 to 70 \GeV.

After applying the above three selections, 169 events are left in the
data while $180.1 \pm 6.3$ events are expected from background. 

\section*{Search for stable charged heavy leptons}

The analysis follows closely the procedure described in
References~\citen{hlep02a}~and~\citen{hlep10} and extends the search to higher masses.
The signal events are characterized by two back-to-back high momentum tracks
in the central tracking detector. Events satisfying the 
following criteria are selected:

\begin{itemize}

\item  The event contains two charged tracks, each with momentum
greater than 5 \GeV\ and polar angle $\rm\vert \cos\theta\vert
< 0.82$. 
\item The acollinearity angle between the 
two tracks is less than $\rm 15^o$.
\end{itemize}

The momentum and acollinearity angle cuts 
reduce the background from two-photon 
produced lepton pairs as well as from  
dilepton annihilation events with a high 
energy photon in the final state.  
16,598 events are selected in the data
samples with 16,715 events expected from Standard Model processes, 
mainly from $\rm e^+ e^- \rightarrow e^+ e^-$.

Stable charged heavy leptons are also expected to be 
highly ionizing in the tracking chamber.
The $dE/dx$ measurement is calibrated with
Bhabha scattering events, and the mean value of the  
$dE/dx$ distribution is normalized to one. 
Its resolution is 0.08 units. 
The ionization energy loss has been studied using 
low energy protons and kaons in our hadronic events up to 
10 units in $dE/dx$ which corresponds to $m_{\rm L}/E_{beam}=$ 0.995.
Figure 5 shows the $dE/dx$ measurements 
of the tracks of data events  
passing the previous cuts, 
as well as the simulated signal 
from pair-production of stable charged heavy leptons.
We select candidate events for which the ionization energy loss for 
each track is between 1.25 and 8.0 units and the product of the track
ionization losses is larger than 2.0 units. The upper cut at
8 units rejects very highly ionizing particles for which 
saturation effects may lead to track reconstruction inefficiencies.

Three candidate events satisfy the selection requirements 
in the data. 
The candidate events were recorded at beam energies of 97.8 
\GeV, 102.7 \GeV and 103.3 \GeV and correspond to 
pair production of stable charged particles of masses
$\rm 82.9\pm 1.8~GeV$ , $\rm 86.5\pm 1.8~GeV$ and $\rm 86.2\pm 2.1~GeV$, 
respectively.
However, no candidate events are observed that are consistent with 
a lepton mass larger than $\rm 90~GeV$.

The background is estimated 
using a calibration sample taken on the 
Z resonance during the 1999 and 2000 data taking periods.
From the number of two track events from this sample passing
all stable heavy lepton selection cuts,
the background is estimated to be $4.1 \pm 1.8$ events
for the entire high energy data sample.

\section*{Results}

No signal for new heavy leptons is observed. The 
number of selected events and expected backgrounds are
in good agreement.
The systematic uncertainty, 
which is mainly due to the uncertainties in the energy 
calibration, the lepton identification efficiency and
purity, and the Monte Carlo statistics, is estimated to be 5\% relative. 
To obtain exclusion limits, the selection efficiency is reduced 
by the total systematic uncertainty. 
These results are combined 
with our earlier L3 data recorded at $\sqrt{s}~=~133-189$ \GeV~\cite{hlep02a,hlep10}.

The selection efficiencies are determined by Monte Carlo simulation. 
For unstable neutral heavy leptons in the mass range
from 90 to 100 \GeV, 
the efficiency is 32\% to 40\% for the electron decay mode,
and 27\% to 35\% for the muon decay mode.
For the tau decay mode, the selection
efficiency ranges from  13\% for a heavy lepton mass of  80 \GeV\ 
to 20\% for a mass of 95 \GeV.
The selection efficiency for unstable charged heavy leptons of mass
100 \GeV\ decaying into light neutrinos is 24\%.
In the case of the charged lepton  decaying into its associated neutral
lepton, the selection efficiency is 5\% to 40\% with increasing mass
difference $\Delta m$.

The selection efficiency for stable charged 
heavy leptons is a function of $m_{\rm L}/E_{beam}$ 
and ranges from 60\% to 70\% over the
range 0.8 to 0.99, including the trigger efficiency.  
For $m_{\rm L}/E_{beam}>0.99$, the efficiency drops due to the
upper cut on track ionization.

Taking into account the luminosities, 
the selection efficiencies and the production
cross sections we calculate the 95\% confidence level, CL, 
lower limits on the masses of heavy leptons~\cite{hlep11}, 
which are shown in Tables 1 and 2. 
The limits for vector heavy leptons
are higher than for the sequential and mirror ones due to their
production cross section.
Figure 6 shows the 95\% CL exclusion contour in the
$m_{\mathrm{L^{\pm}}}-m_{\mathrm{L^0}}$ mass plane for decays of the
charged lepton into its associated neutral lepton. The lower
limit on the mass of the charged heavy lepton is shown in Table 2 for a mass
difference greater than 15 \GeV.
These limits include the full LEP data set and improve upon and supersede
our previous results.

\begin{table}[htb]
\begin{center}
\caption{95\% CL lower mass limits in GeV for 
pair-produced neutral heavy leptons.}
\label{tab:lnlep}
\vspace{0.4cm}
\begin{tabular}{|c|c|c|c|}
\hline
 Decay mode & Model & Dirac & Majorana \\
\hline
                        & Sequential & 101.3 &  89.5 \\
L$^0 \rightarrow $eW    & Vector     & 102.6 &  $-$ \\
                        & Mirror     & 100.8 &  89.5 \\
\hline
                        & Sequential & 101.5 &  90.7 \\
L$^0 \rightarrow \mu$W  & Vector     & 102.7 &  $-$ \\
                        & Mirror     & 101.0 &  90.7 \\
\hline
                        & Sequential & 90.3 &  80.5 \\
L$^0 \rightarrow \tau$W & Vector     & 99.3 &  $-$ \\
                        & Mirror     & 90.3 &  80.5 \\
\hline
\end{tabular}
\end{center}
\end{table}

\begin{table}[htb]
\begin{center}
\caption{95\% CL lower mass limits in GeV 
for pair-produced charged heavy leptons.}
\label{tab:lclep}
\vspace{0.4cm}
\begin{tabular}{|c|c|c|}
\hline
 Decay mode & Model &  \\
\hline
                        & Sequential & 100.8 \\
L$^\pm \rightarrow \nu$W  & Vector     & 101.2 \\
                        & Mirror     & 100.5 \\
\hline
                        & Sequential & 101.9 \\
L$^\pm \rightarrow$ L$^0$W  & Vector     & 102.1 \\
                        & Mirror     & 101.9 \\
\hline
                        & Sequential & 102.6 \\
Stable                  & Vector     & 102.6 \\
                        & Mirror     & 102.6 \\
\hline
\end{tabular}
\end{center}
\end{table}

\section*{Acknowledgements}

We wish to express our gratitude to the CERN accelerator divisions for the 
excellent performance of the LEP machine. We acknowledge with appreciation 
the effort of the engineers, technicians and support staff who have
participated in the construction and maintenance of this experiment.

%
%

\vfill

\newpage



\newpage
\typeout{   }     
\typeout{Using author list for paper 237 -- ? }
\typeout{$Modified: Fri Jan 26 2001 by smele $}
\typeout{!!!!  This should only be used with document option a4p!!!!}
\typeout{   }
%
%
%
%
%
%

\newcount\tutecount  \tutecount=0
\def\tutenum#1{\global\advance\tutecount by 1 \xdef#1{\the\tutecount}}
\def\tute#1{$^{#1}$}
\tutenum\aachen            
\tutenum\nikhef            
\tutenum\mich              
\tutenum\lapp              
\tutenum\basel             
\tutenum\lsu               
\tutenum\beijing           
\tutenum\berlin            
\tutenum\bologna           
\tutenum\tata              
\tutenum\ne                
\tutenum\bucharest         
\tutenum\budapest          
\tutenum\mit               
\tutenum\panjab            
\tutenum\debrecen          
\tutenum\florence          
\tutenum\cern              
\tutenum\wl                
\tutenum\geneva            
\tutenum\hefei             
\tutenum\lausanne          
\tutenum\lyon              
\tutenum\madrid            
\tutenum\florida           
\tutenum\milan             
\tutenum\moscow            
\tutenum\naples            
\tutenum\cyprus            
\tutenum\nymegen           
\tutenum\caltech           
\tutenum\perugia           
\tutenum\peters            
\tutenum\cmu               
\tutenum\potenza           
\tutenum\prince            
\tutenum\riverside         
\tutenum\rome              
\tutenum\salerno           
\tutenum\ucsd              
\tutenum\sofia             
\tutenum\korea             
\tutenum\utrecht           
\tutenum\purdue            
\tutenum\psinst            
\tutenum\zeuthen           
\tutenum\eth               
\tutenum\hamburg           
\tutenum\taiwan            
\tutenum\tsinghua          

{
\parskip=0pt
\noindent
{\bf The L3 Collaboration:}
\ifx\selectfont\undefined
 \baselineskip=10.8pt
 \baselineskip\baselinestretch\baselineskip
 \normalbaselineskip\baselineskip
 \ixpt
\else
 \fontsize{9}{10.8pt}\selectfont
\fi
\medskip
\tolerance=10000
\hbadness=5000
\raggedright
\hsize=162truemm\hoffset=0mm
\def\r{\rlap,}
\noindent

P.Achard\r\tute\geneva\ 
O.Adriani\r\tute{\florence}\ 
M.Aguilar-Benitez\r\tute\madrid\ 
J.Alcaraz\r\tute{\madrid,\cern}\ 
G.Alemanni\r\tute\lausanne\
J.Allaby\r\tute\cern\
A.Aloisio\r\tute\naples\ 
M.G.Alviggi\r\tute\naples\
H.Anderhub\r\tute\eth\ 
V.P.Andreev\r\tute{\lsu,\peters}\
% ...
\newpage
\begin{list}{A}{\itemsep=0pt plus 0pt minus 0pt\parsep=0pt plus 0pt minus 0pt
                \topsep=0pt plus 0pt minus 0pt}
\item[\aachen]
 I. Physikalisches Institut, RWTH, D-52056 Aachen, FRG$^{\S}$\\
 III. Physikalisches Institut, RWTH, D-52056 Aachen, FRG$^{\S}$
\item[\nikhef] National Institute for High Energy Physics, NIKHEF, 
     and University of Amsterdam, NL-1009 DB Amsterdam, The Netherlands
\item[\mich] University of Michigan, Ann Arbor, MI 48109, USA
\item[\lapp] Laboratoire d'Annecy-le-Vieux de Physique des Particules, 
     LAPP,IN2P3-CNRS, BP 110, F-74941 Annecy-le-Vieux CEDEX, France
\item[\basel] Institute of Physics, University of Basel, CH-4056 Basel,
     Switzerland
\item[\lsu] Louisiana State University, Baton Rouge, LA 70803, USA
\item[\beijing] Institute of High Energy Physics, IHEP, 
  100039 Beijing, China$^{\triangle}$ 
\item[\berlin] Humboldt University, D-10099 Berlin, FRG$^{\S}$
\item[\bologna] University of Bologna and INFN-Sezione di Bologna, 
     I-40126 Bologna, Italy
\item[\tata] Tata Institute of Fundamental Research, Mumbai (Bombay) 400 005, India
\item[\ne] Northeastern University, Boston, MA 02115, USA
\item[\bucharest] Institute of Atomic Physics and University of Bucharest,
     R-76900 Bucharest, Romania
\item[\budapest] Central Research Institute for Physics of the 
     Hungarian Academy of Sciences, H-1525 Budapest 114, Hungary$^{\ddag}$
\item[\mit] Massachusetts Institute of Technology, Cambridge, MA 02139, USA
\item[\panjab] Panjab University, Chandigarh 160 014, India.
\item[\debrecen] KLTE-ATOMKI, H-4010 Debrecen, Hungary$^\P$
\item[\florence] INFN Sezione di Firenze and University of Florence, 
     I-50125 Florence, Italy
\item[\cern] European Laboratory for Particle Physics, CERN, 
     CH-1211 Geneva 23, Switzerland
\item[\wl] World Laboratory, FBLJA  Project, CH-1211 Geneva 23, Switzerland
\item[\geneva] University of Geneva, CH-1211 Geneva 4, Switzerland
\item[\hefei] Chinese University of Science and Technology, USTC,
      Hefei, Anhui 230 029, China$^{\triangle}$
\item[\lausanne] University of Lausanne, CH-1015 Lausanne, Switzerland
\item[\lyon] Institut de Physique Nucl\'eaire de Lyon, 
     IN2P3-CNRS,Universit\'e Claude Bernard, 
     F-69622 Villeurbanne, France
\item[\madrid] Centro de Investigaciones Energ{\'e}ticas, 
     Medioambientales y Tecnolog{\'\i}cas, CIEMAT, E-28040 Madrid,
     Spain${\flat}$ 
\item[\florida] Florida Institute of Technology, Melbourne, FL 32901, USA
\item[\milan] INFN-Sezione di Milano, I-20133 Milan, Italy
\item[\moscow] Institute of Theoretical and Experimental Physics, ITEP, 
     Moscow, Russia
\item[\naples] INFN-Sezione di Napoli and University of Naples, 
     I-80125 Naples, Italy
\item[\cyprus] Department of Physics, University of Cyprus,
     Nicosia, Cyprus
\item[\nymegen] University of Nijmegen and NIKHEF, 
     NL-6525 ED Nijmegen, The Netherlands
\item[\caltech] California Institute of Technology, Pasadena, CA 91125, USA
\item[\perugia] INFN-Sezione di Perugia and Universit\`a Degli 
     Studi di Perugia, I-06100 Perugia, Italy   
\item[\peters] Nuclear Physics Institute, St. Petersburg, Russia
\item[\cmu] Carnegie Mellon University, Pittsburgh, PA 15213, USA
\item[\potenza] INFN-Sezione di Napoli and University of Potenza, 
     I-85100 Potenza, Italy
\item[\prince] Princeton University, Princeton, NJ 08544, USA
\item[\riverside] University of Californa, Riverside, CA 92521, USA
\item[\rome] INFN-Sezione di Roma and University of Rome, ``La Sapienza",
     I-00185 Rome, Italy
\item[\salerno] University and INFN, Salerno, I-84100 Salerno, Italy
\item[\ucsd] University of California, San Diego, CA 92093, USA
\item[\sofia] Bulgarian Academy of Sciences, Central Lab.~of 
     Mechatronics and Instrumentation, BU-1113 Sofia, Bulgaria
\item[\korea]  The Center for High Energy Physics, 
     Kyungpook National University, 702-701 Taegu, Republic of Korea
\item[\utrecht] Utrecht University and NIKHEF, NL-3584 CB Utrecht, 
     The Netherlands
\item[\purdue] Purdue University, West Lafayette, IN 47907, USA
\item[\psinst] Paul Scherrer Institut, PSI, CH-5232 Villigen, Switzerland
\item[\zeuthen] DESY, D-15738 Zeuthen, 
     FRG
\item[\eth] Eidgen\"ossische Technische Hochschule, ETH Z\"urich,
     CH-8093 Z\"urich, Switzerland
\item[\hamburg] University of Hamburg, D-22761 Hamburg, FRG
\item[\taiwan] National Central University, Chung-Li, Taiwan, China
\item[\tsinghua] Department of Physics, National Tsing Hua University,
      Taiwan, China
\item[\S]  Supported by the German Bundesministerium 
        f\"ur Bildung, Wissenschaft, Forschung und Technologie
\item[\ddag] Supported by the Hungarian OTKA fund under contract
numbers T019181, F023259 and T024011.
\item[\P] Also supported by the Hungarian OTKA fund under contract
  number T026178.
\item[$\flat$] Supported also by the Comisi\'on Interministerial de Ciencia y 
        Tecnolog{\'\i}a.
\item[$\sharp$] Also supported by CONICET and Universidad Nacional de La Plata,
        CC 67, 1900 La Plata, Argentina.
\item[$\triangle$] Supported by the National Natural Science
  Foundation of China.
\end{list}
}
\vfill


\newpage


\begin{figure}[p]
\begin{center}
\mbox{\epsfysize=16cm\epsffile{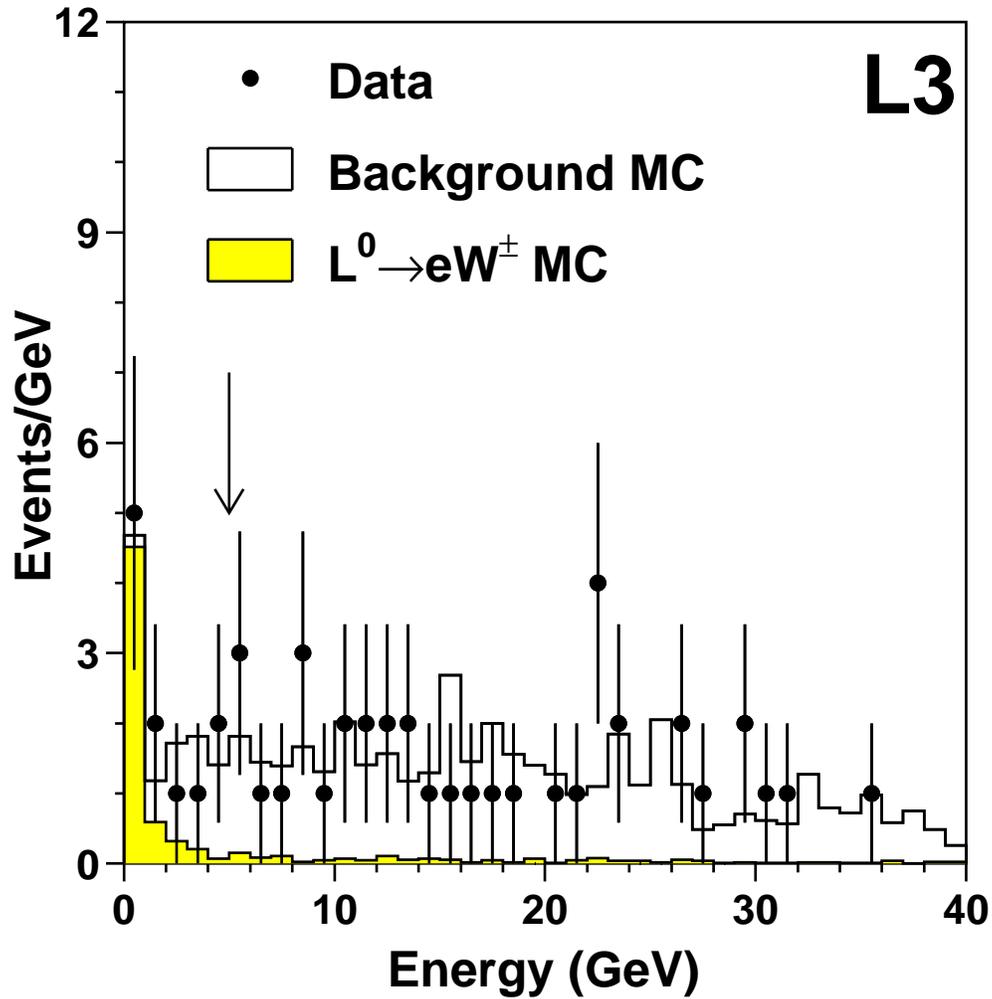}}
\end{center}
\caption{The distribution of the 
energy in a 30$^{\circ}$ cone around the second most energetic electron
candidate. The points are the data, the solid histogram
is the background Monte Carlo. The shaded histogram represents the simulated
signal for e$^+ $e$^- \rar \mathrm{L^0}\overline{\mathrm{L^0}}$
for a sequential Dirac neutral lepton mass of 101 \GeV.
Both histograms are normalized to the luminosity of the data.
The arrow indicates the value of the applied cut;
all other selection cuts are applied.
}
\label {isolation}
\end{figure}

\begin{figure}[p]
\begin{center}
\mbox{\epsfysize=16cm\epsffile{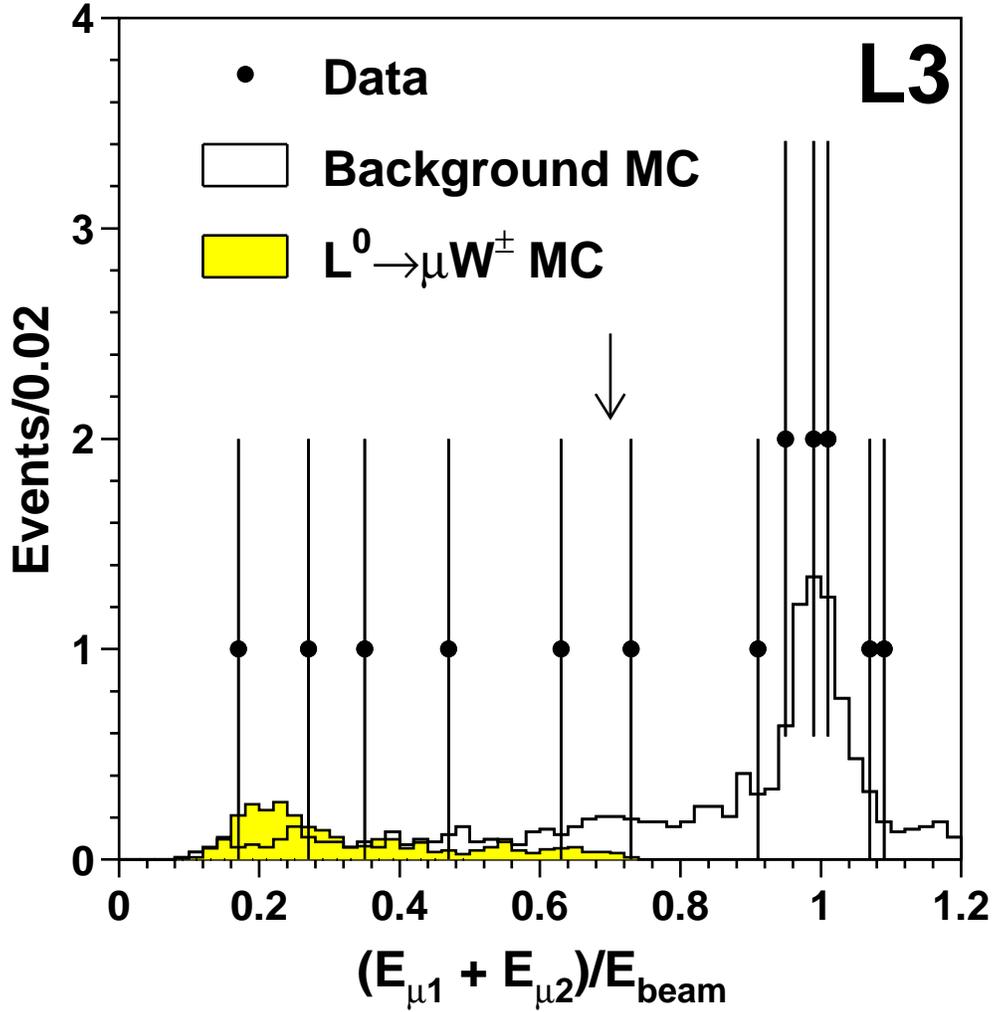}}
\end{center}
\caption{The distribution of the sum of the 
muon energies, $E_{\mu 1}+E_{\mu 2}$, normalized to
the beam enery.
The points are the data, the solid histogram
is the background Monte Carlo. The shaded histogram represents the simulated
signal for e$^+ $e$^- \rar \mathrm{L^0}\overline{\mathrm{L^0}}$
for a sequential Dirac neutral lepton mass equal to 101 \GeV.
Both histograms are normalized to the luminosity of the data.
The arrow indicates the value of the applied cut;
all other selection cuts are applied.
}
\label {sum_mu}
\end{figure}

\begin{figure}[p]
\begin{center}
\mbox{\epsfysize=16cm\epsffile{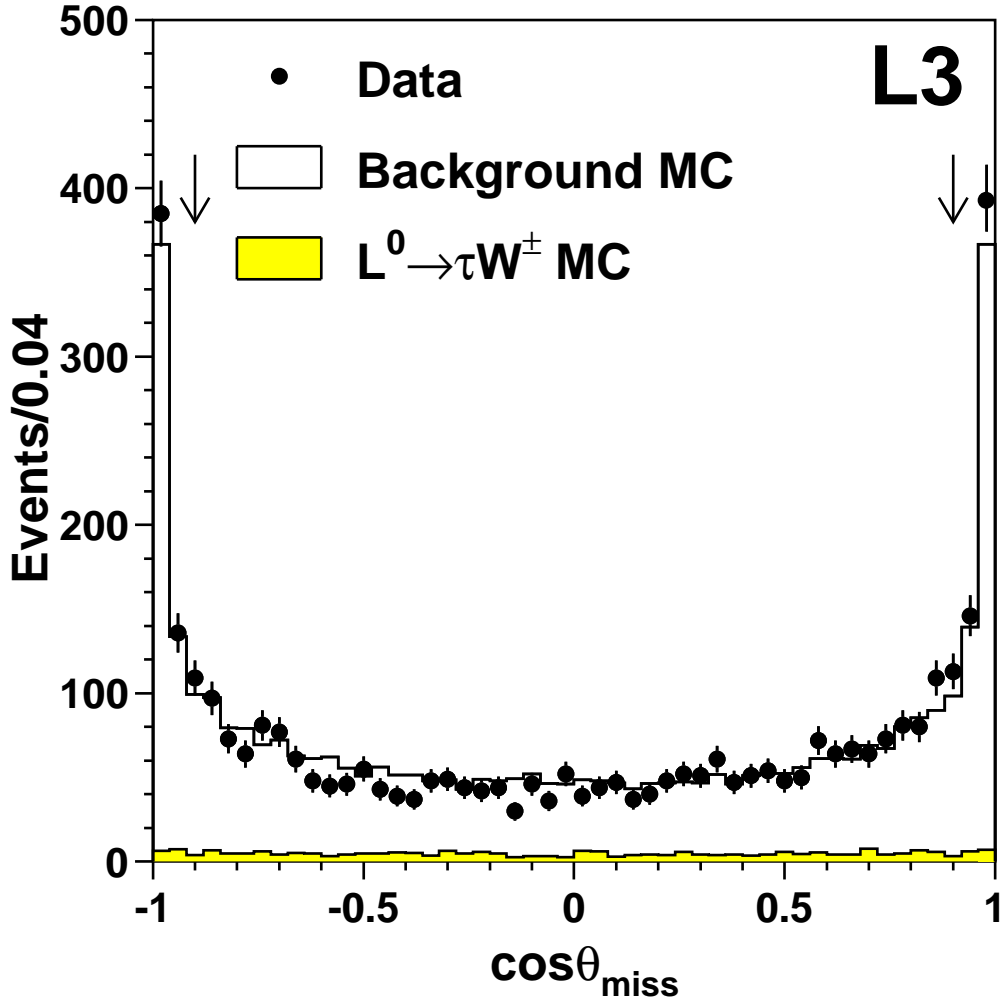}}
\end{center}
\caption{The distribution of the cosine of the polar angle of the 
missing momentum, $\cos\theta_{miss}$.
The points are the data, the solid histogram
is the background Monte Carlo. The shaded histogram represents the simulated
signal for e$^+ $e$^- \rar \mathrm{L^0}\overline{\mathrm{L^0}}$
for a sequential Dirac neutral lepton mass equal to 80 \GeV.
The normalization for the signal Monte Carlo is scaled by a factor
of 2 for better visibility.
The arrows indicate the values of the applied cut.
}
\label {tau}
\end{figure}

\begin{figure}[p]
\begin{center}
\mbox{\epsfysize=16cm\epsffile{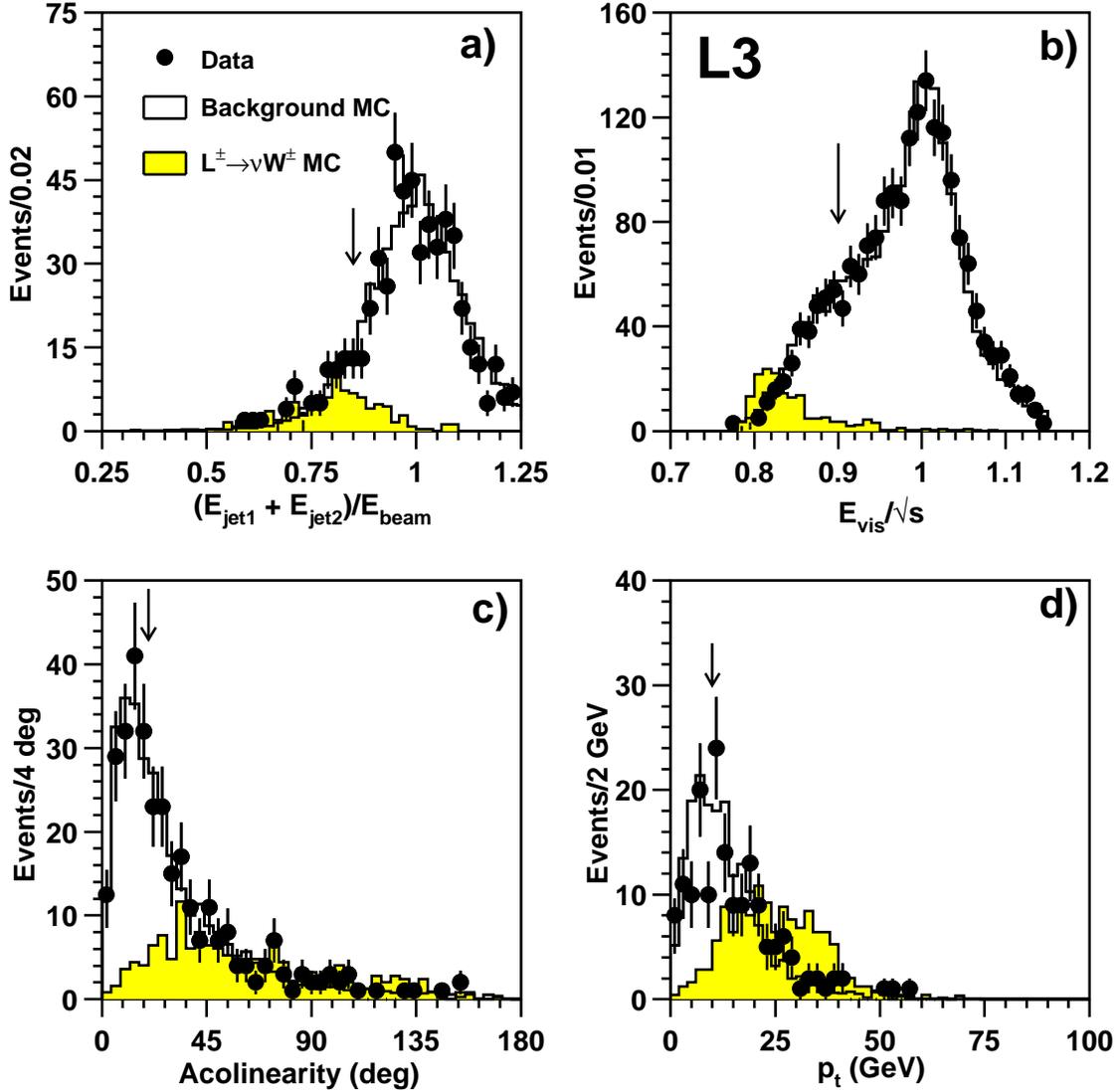}}
\end{center}
\caption{The distribution of 
a) the sum, $E_{jet_1} + E_{jet_2}$, of the energies of the hadronic jets
normalized to the beam energy,
b) the normalized visible energy in the event,
c) the acollinearity angle of the two W candidates,
d) the total transverse momentum, $p_t$.
The points are the data, the solid histogram
is the background Monte Carlo. 
The shaded histogram represents the simulated
signal for e$^+ $e$^- \rar \mathrm{L^+} \mathrm{L^-}$
for a sequential charged lepton mass equal to 95 \GeV.
The normalization for the signal Monte Carlo is scaled by a factor
of 2 for better visibility.
The arrows indicate the values of the applied cuts
after all the other selection requirements.
}
\label {4fig}
\end{figure}

\begin{figure}[p]
\begin{center}
\mbox{\epsfysize=16cm\epsffile{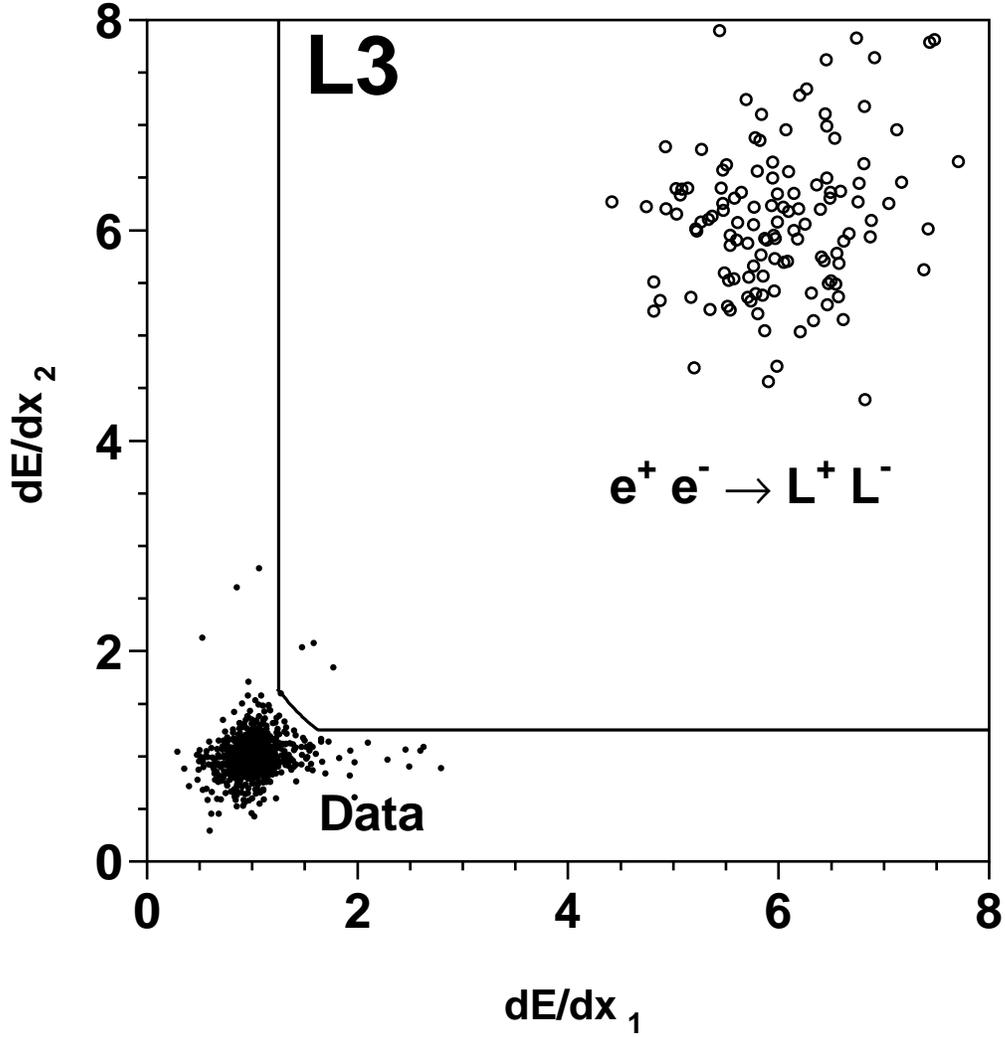}}
\end{center}
\caption{The two dimensional distribution of the 
ionization energy loss of the most energetic track, $dE/dx_1$,
and the least energetic track, $dE/dx_2$.
The solid circles represent the data and the open circles 
represent
the simulated signal for e$^+ $e$^- \rar \mathrm{L^+} \mathrm{L^-}$  
for a lepton mass of $0.985\times E_{beam}$. 
The signal normalization is arbitrary.
The lines indicate the values of the applied cut. 
}
\label {scat9900}
\end{figure}

\begin{figure}[p]
\begin{center}
\mbox{\epsfysize=16cm\epsffile{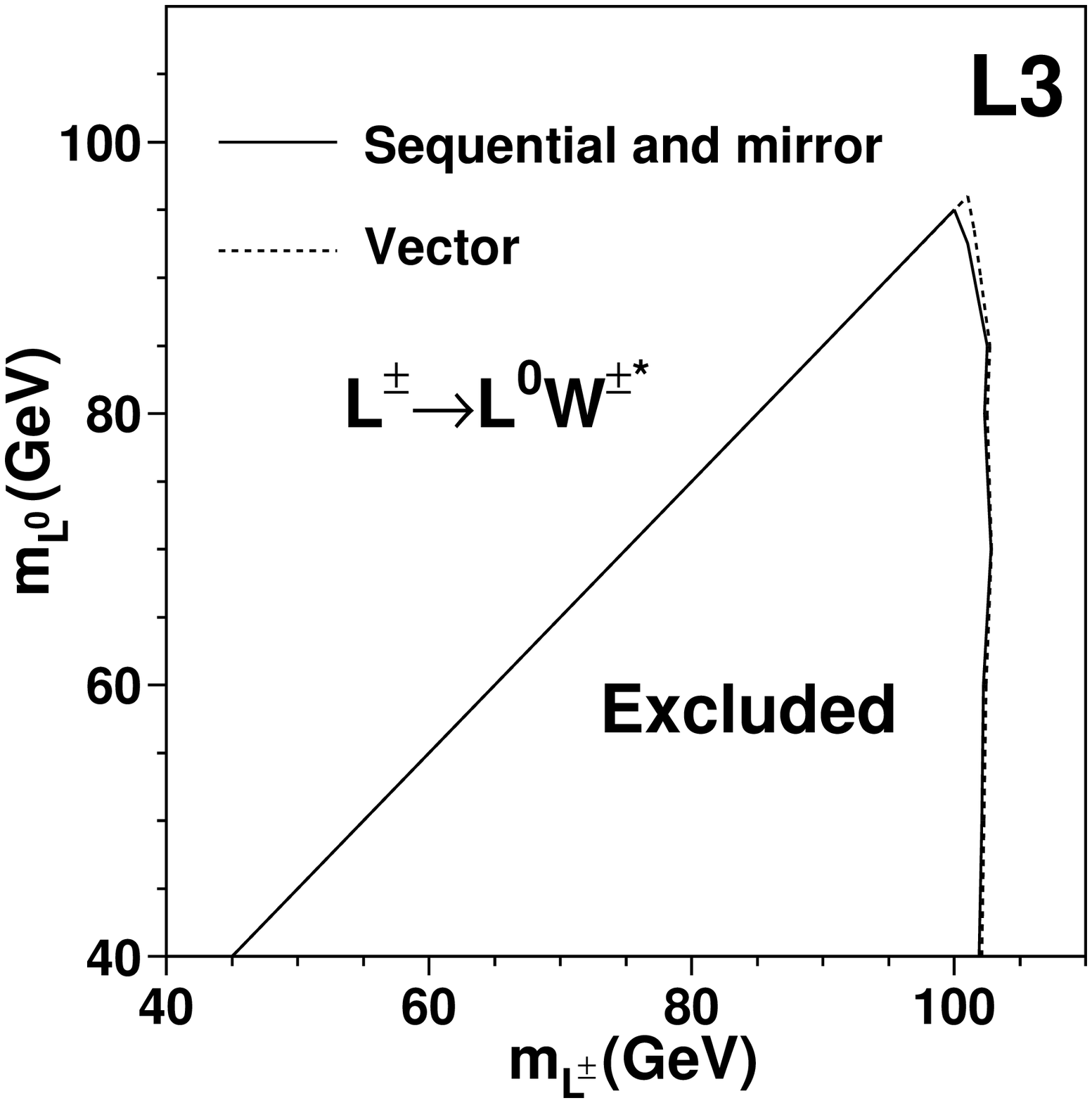}}
\end{center}
\caption{The 95\% CL limit in the 
$m_{\mathrm{L^{\pm}}}-m_{\mathrm{L^0}}$ plane for
the case of a charged heavy lepton decaying into a stable
neutral heavy lepton. The solid and dashed lines represent 
respectively the sequential and mirror, and vector types of heavy
lepton.
}
\label {clcn}
\end{figure}

\begin{thebibliography}{10}                                                     
\bibitem{hlep00}
S.~L. Glashow, Nucl. Phys. {\bf 22} (1961) 579;\newline
S. Weinberg, Phys. Rev. Lett. {\bf 19} (1967) 1264;\newline
A. Salam, ``Weak and Electromagnetic Interactions'', in {\it Elementary
Particle Theory}, edited by N. Svartholm, page 367, Stockholm, 1968,
Almqvist and Wiksell.

\bibitem{hlep01}
For a review see:\newline 
D. London, in {\it Precision Tests of the Standard Model}, 
ed. P. Langacker, World Scientific, Singapore (1995);\newline
A. Djouadi, J. Ng and T.G. Rizzo,
in {\it Electroweak Symmetry Breaking and
New Physics at the TeV Scale}, eds. 
T.~Barklow {\em et al.},
World Scientific, Singapore (1997).

\bibitem{hlep02}
ALEPH Collaboration, D. Decamp {\em et al.}, Phys. Rep. {\bf 216} (1992) 
253;\newline
DELPHI Collaboration, P. Abreu {\em et al.}, Phys. Lett. {\bf B 274} (1992)
230;\newline
L3 Collaboration, O. Adriani {\em et al.}, Phys. Rep. {\bf 236} (1993)
1;\newline
OPAL Collaboration, M.Z. Akrawy {\em et al.}, Phys. Lett. {\bf B 240} (1990)
250;\newline
OPAL Collaboration, M.Z. Akrawy {\em et al.}, Phys. Lett. {\bf B 247} (1990)
448;\newline
OPAL Collaboration, M.Z. Akrawy {\em et al.}, Phys. Lett. {\bf B 252} (1990)
290;\newline
MARK II Collaboration, G.S. Abrams {\em et al.}, Phys. Rev. Lett. {\bf 63}
(1989) 2447;\newline
MARK II Collaboration, E. Soderstrom {\em et al.}, Phys. Rev. Lett. {\bf 64}
(1990) 2980. 

\bibitem{hlep02a}
L3 Collaboration, M. Acciarri {\em et al.}, Phys. Lett. {\bf B 462} (1999)
354.

\bibitem{hlep03}
ALEPH Collaboration, D. Buskulic {\em et al.}, Phys. Lett. {\bf B 384} (1996) 
439;\newline
ALEPH Collaboration, R. Barate {\em et al.}, Phys. Lett. {\bf B 405} (1997)
379;\newline
DELPHI Collaboration, P. Abreu {\em et al.}, Eur. Phys. J. {\bf C 8} (1999) 
41;\newline
DELPHI Collaboration, P. Abreu {\em et al.}, Phys. Lett. {\bf B 478} (2000) 
65;\newline
OPAL Collaboration, K. Ackerstaff {\em et al.}, Phys. Lett. {\bf B 433} (1998)
195;\newline
OPAL Collaboration, K. Ackerstaff {\em et al.}, Eur. Phys. J. {\bf C 14} (2000) 73.

\bibitem{hlep04}                                                           
L3 Collaboration, B. Adeva {\em et al.}, Nucl. Instr. Meth. {\bf A 289}
(1990) 35;\newline
M. Acciarri {\em et al.}, Nucl. Instr. Meth. {\bf A 351}
(1994) 300;\newline 
M. Chemarin {\em et al.}, Nucl. Instr. Meth. {\bf A 349} (1994) 345;\newline
M. Adam {\em et al.}, Nucl. Instr. Meth. {\bf A 383} (1996) 342;\newline
G. Basti {\em et al.}, Nucl. Instr. Meth. {\bf A 374} (1996) 293.

\bibitem{hlep041}
N. Evans, Phys. Lett. {\bf B 340} (1994) 81;\newline
P. Bamert and C.P. Burgess, Z. Phys. {\bf C 66} (1995) 495;\newline
T. Inami {\em et al.}, Mod. Phys. Lett. {\bf A 10} (1995) 1471;\newline
A. Masiero {\em et al.}, Phys. Lett. {\bf B 355} (1995) 329;\newline
V.~A. Novikov, 
{\em et al.}, Mod. Phys. Lett. {\bf A 10}
(1995) 1915; Erratum - ibid. {\bf A 11} (1996) 698;\newline
The Particle Data Group, C. Caso {\em et al.}, Eur. Phys. J. {\bf C 3}
(2000) 95.

\bibitem{hlep042}
For a review, see J. Hewett and T.G. Rizzo, Phys. Rep. {\bf 183}, (1989) 193.

\bibitem{hlep043}
M. Maalampi and M. Roos, Phys. Rep. {\bf 186} (1990) 53.

\bibitem{hlep051}
S.L. Glashow, J. Iliopoulos and L. Maiani, Phys. Rev. {\bf D 2} (1970) 1285.

\bibitem{hlep055}
M. Gronau, C. Leung and J. Rosner, Phys. Rev. {\bf D 29} (1984) 2539.

\bibitem{hlep06}
S. Jadach and J. K\"uhn, TIPTOP Monte Carlo, Preprint MPI-PAE/PTh 64/86.

\bibitem{hlep060} KK2F version 4.12 is used:
S. Jadach, B.F.L. Ward and Z. W\c{a}s, Comp. Phys. Comm. {\bf 130} (2000) 260.

\bibitem{hlep07} PYTHIA version 5.772 and JETSET version 7.409 are used: 
T.~Sj\"ostrand, Comp. Phys. Comm. {\bf 82} (1994) 74.

\bibitem{hlep070} BHWIDE version 1.01 is used:
S. Jadach {\em et al.}, Phys. Lett. {\bf D 390} (1997) 298.

\bibitem{hlep071} KORALZ version 4.02 is used:
S. Jadach, J. K\"uhn and Z. W\c{a}s, Comp. Phys. Comm. {\bf 64} (1991) 275;\newline
S. Jadach, B.F.L.Ward and Z. W\c{a}s, Comp. Phys. Comm. {\bf 66} (1991) 367.

\bibitem{hlep072} KORALW version 1.33 is used:
S. Jadach {\em et al.,} Comp. Phys. Comm. {\bf 94} (1996) 216;\\
S. Jadach {\em et al.,} Phys. Lett. {\bf B 372} (1996) 289. 

\bibitem{hlep08} PHOJET version 1.05 is used:
R. Engel, Z. Phys. {\bf C 66} (1993) 1657;\\
R. Engel and J. Ranft, Phys. Rev. {\bf D 54} (1996) 4244.

\bibitem{hlep081}
F.A. Berends, P.H. Daverveldt and R. Kleiss, Nucl. Phys. {\bf B 253} (1985) 441.

\bibitem{hlep082}
F.A. Berends, R. Kleiss and R. Pittau, Comp. Phys. Comm. {\bf 85} (1995) 437.

\bibitem{hlep09}
R. Brun et al., GEANT 3.15 preprint CERN DD/EE/84-1 (1984), revised 1987;
H. Fesefeldt, RWTH Aachen report PITHA 85/2 (1985).

\bibitem{hlep091}
S. Catani {\em et al.}, Phys. Lett. {\bf B 269} (1991) 432;\newline  
S. Bethke {\em et al.}, Nucl. Phys. {\bf B 370} (1992) 310.

\bibitem{charg}
L3 Collab., M.~Acciarri {\it et al.}, Phys. Lett. {\bf 472} (2000) 420.

\bibitem{hlep10}
L3 Collab., M.~Acciarri {\it et al.}, Phys. Lett. {\bf 462} (1999) 354.

\bibitem{hlep11}
V.F.~Obraztsov, Nucl. Instr. Meth. {\bf A 316} (1992) 388.

\end{thebibliography}
\end{document}